\documentclass[12]{iopart}

\expandafter\let\csname equation*\endcsname\relax
\expandafter\let\csname endequation*\endcsname\relax

\usepackage{graphicx}
\usepackage{amssymb}
\usepackage{amsmath}
\usepackage{bm}
\usepackage{url}
\usepackage{epstopdf}
\usepackage{iopams}

\begin{document}

\title[Single beam, Two-Colour Free Electron Laser with Wide Frequency Separation]{Two-Colour Free Electron Laser with Wide Frequency Separation using a Single Monoenergetic Electron Beam}
\author{L.T. Campbell$^{1,2,3,4}$, B.W.J. M$^{\mathrm c}$Neil$^3$ and S. Reiche$^5$}
\address{1 Center for Free-Electron Laser Science, Notkestrasse 85, Hamburg, Germany} 
\address{2 Institut f\"{u}r Experimentalphysik, Universit\"{a}t Hamburg, Hamburg, Germany}
\address{3 SUPA, Department of Physics, University of Strathclyde, Glasgow, UK}
\address{4 ASTeC, STFC Daresbury Laboratory and Cockcroft Institute, Warrington WA4 4AD, United Kingdom}
\address{5 Paul Scherrer Institute, Villigen PSI, Switzerland}
\eads{\mailto{lawrence.campbell@strath.ac.uk}, \mailto{b.w.j.mcneil@strath.ac.uk}, \mailto{sven.reiche@psi.ch}}
\submitto{\NJP}

\maketitle

\begin{abstract}
Studies of a broad bandwidth, two-colour FEL amplifier using one monoenergetic electron beam are presented. The two-colour FEL interaction is achieved using a series of undulator modules alternately tuned to two well-separated resonant frequencies. Using the broad bandwidth FEL simulation code Puffin, the electron beam is shown to  bunch strongly and simultaneously at the two resonant frequencies.  Electron bunching components are also generated at the sum and difference of the resonant frequencies. 
\end{abstract}

\section{Introduction}

X-ray Free Electron Lasers are now operating~\cite{lcls,sacla} or under construction around the world~\cite{nprev}. These provide sources of high brightness, monochromatic X-rays that are driving forward many new scientific discoveries. There is now also increasing user interest in the simultaneous delivery of such radiation at two distinct colours, either simultaneously, or with a short time-delay in a pump-probe architecture. 

Two colour operation of an FEL was first reported with a relatively low colour contrast in an infra-red oscillator configuration~\cite{dino}, with greater colour contrasts also possible~\cite{Krishnagopal}. 
More recently, two colour operation has been achieved from the UV into the X-ray using different configurations of an FEL amplifier. For example, two-colour operation is being investigated using an electron beam composed of `beamlets' of more than one energy~\cite{2beam,2beam2,3beam2}. These beamlets, typically of energy separation greater than the FEL amplification bandwidth, propagate through identical undulator modules and lase independently and simultaneously. If the higher energy beamlet emits at an harmonic of the lower energy, then coupling between the beamlets that is favourable to shorter wavelength generation can occur~\cite{2beamhar}. 

An alternative approach has been used in~\cite{slac,2colSACLA} using a single monoenergetic electron beam propagating sequentially through two undulator modules each tuned to the two different colours.  This method can generate two-colour radiation with a frequency separation of $\gtrsim 30\%$~\cite{2colSACLA}.  In~\cite{gainMod}, a frequency separation of $\sim 4\%$ was demonstrated using an alternating series of undulator modules tuned to the two colours. This method generates radiation frequencies with a temporal overlap. In addition to the beating effects of the two resonant colours~\cite{3beam2}, a modal spectral structure~\cite{mlsase} may arise due to the temporal shifts of each colour as it propagates non-resonantly over the electrons in alternate undulator modules.


Averaged FEL simulation codes such as~\cite{MEDUSSA,GENESIS,FAST,GINGER} that make the Slowly Varying Envelope Approximation (SVEA)~\cite{SVEA}, can readily model 2-colour FEL interactions when the radiation wavelengths are close to a fundamental frequency $\omega_1$ and e.g. its third harmonic $\omega_3 = 3\omega_1$. This requires the radiation to be described by two distinct computational fields as the field sampling at any resonant radiation frequency $\omega_r$ means the frequency range able to be sampled is limited by the Nyquist condition to $\omega_r/2<\omega<3\omega_r/2$. Furthermore, the accuracy of the model decreases the further away from the resonant frequency.   For a fundamental and e.g. its 3rd harmonic field, the sampling can be over a common set of `slices,' each an integer number of fundamental wavelengths in length. 

However, for cases where the two frequencies in a 2-colour FEL  are strongly non-harmonic and well separated, e.g. where $\omega_2 = 0.417 \times \omega_1$,  modeling of two computational fields using the averaged SVEA approximation becomes problematic -  integer wavelength long slices of the two fields cannot be coincident and so when radiation propagation effects are modeled, electrons within these different length slices are driven asymmetrically in phase space leading to unphysical instabilities.

The unaveraged free electron laser simulation code Puffin~\cite{POP} does not make the SVEA approximation and is therefore not restricted to the limited bandwidth of conventional averaged FEL codes. We therefore define such unaveraged simulation codes as being `broad-bandwidth'. Furthermore, because Puffin  also models the electron beam dynamics without  averaging,  the electrons are not confined to localised `slices' of width the fundamental  radiation wavelength, and  a realistic electron beam interaction can occur over a broad bandwidth of radiation wavelengths. Puffin (along with other unaveraged codes e.g.~\cite{muffin,femfel,hooker,maroli}) is then ideally suited to simulating and investigating the physics of multi-color FELs.

In what follows, the Puffin code was  modified to allow for undulator modules of different undulator parameters $\bar{a}_w$. This code was then used to simulate a high-gain 2-colour FEL amplifier, with undulator modules alternately tuned to well separated, non-harmonic wavelengths. It is not immediately clear how such an interaction would be expected to progress and in particular how the electron phase space would develop when interacting resonantly with two distinct radiation wavelengths. The interaction is therefore first seeded, rather than starting from noise, to allow a clearer picture of the electron beam and radiation evolution in the 2-colour interaction. An example is then given of an unseeded, 2-colour SASE FEL interaction.  

\section{Mathematical Model \label{sectMath}}
The resonant wavelength of the FEL interaction is given by~\cite{nprev}:
\begin{align}
\lambda=\frac{\lambda_w}{2\gamma_0^2}\left(1+\bar{a}_w^2\right),
\end{align}
where $\bar{a}_w \propto B_w\lambda_w$ is the  rms undulator parameter, $B_w$ is the rms undulator magnetic field strength,  $\lambda_w$ is the undulator period and $\gamma_0$ is the initial electron beam Lorentz factor. Alternating the $\bar{a}_w$ of successive undulator modules, such as used in~\cite{gainMod}, generates the two resonant colours in the FEL interaction for a single monoenergetic electron beam. A variable undulator parameter is defined as $\bar{a}_{w}(\bar{z}) = \alpha(\bar{z})\bar{a}_{w1}$, where $\bar{a}_{w1}$ is a constant initial value. 

Puffin uses a system of equations which utilises scaled variables developed in~\cite{bnp}. These dimensionless variables are scaled with respect to the FEL parameter, defined as:
\begin{align}
\label{rho}
\rho=\frac{1}{\gamma_0}\left(\frac{\bar{a}_w \omega_p}{4ck_w}\right)^{2/3},
\end{align}
 and $k_w = 2\pi / \lambda_w$. To model alternate undulator modules of different tuning, the modules are tuned by altering the rms undulator magnetic field only. A change in the tuning therefore also changes the FEL parameter, and so it is convenient to redefine the scaling used in Puffin to account for the different undulator tunings. The variable FEL parameter is then defined as: 
\begin{align}
\rho = \alpha^{2/3} \rho_1, \label{rhon}
\end{align}
where $\rho_1$ is a constant using $\bar{a}_{w1}$ and the $\bar{z}$ dependence of $\alpha$ is understood. The electron beam focusing factor $f$ of~\cite{POP}, is made variable to maintain a constant beam radius throughout different module tunings so that $f = \alpha f_1$. The Puffin equations can then be re-scaled with respect to the constant $\rho_1$ and $\bar{a}_{w1}$, making $\alpha$ explicit. The resulting equations are identical in form to equations $(31-36)$ of~\cite{POP}, with the exception of two multiplicative factors of $\alpha$ in equations $(32)$ and $(33)$, which describe the evolution of the scaled transverse and longitudinal electron momentum $\bar{p}_{\bot}$ and $p_2$, respectively. Ignoring the explicit scaling of the focusing terms, these become:
\begin{align}
\frac{d\bar{p}_{\bot j}}{d\bar{z}} & = \frac{1}{2\rho_1}\left[ i \alpha U^* -  \frac{\eta p_{2j}}{f^2\bar{k}_{\beta}^2} A_{\bot j} \right] \label{pperpdotgen}   
\end{align}

\begin{align}
\frac{dp_{2j}}{d\bar{z}} = \frac{2\rho_1}{u^2 \eta} & L_{j}^2 \biggl[ \eta p_{2j} (A^{\ast}_{\bot j}\bar{p}_{\bot j} + c.c.)\nonumber \\  
& -i \alpha (1+\eta p_{2j})f^2 \bar{k}_{\beta}^2(U\bar{p}_{\bot j} -c.c.)\biggr], \label{Qeqgen} 
\end{align}
where the scaled interaction length is $\bar{z}=z/l_{g1}$ and $l_{g1}=\lambda_w/4\pi\rho_1$ is the gain length of the $\alpha=1$ interaction.

The system of equations obtained is general, and $\alpha$ can be made any function of distance through the interaction as required. It could, for example, be used to taper the undulator in Puffin, or to provide any number of differently tuned undulator modules. For the purposes of this paper, $\alpha$ is varied between  two values $\alpha_{1,2}$, corresponding to the two colours of FEL radiation, $\omega_{1,2}$.

In the 2-colour FEL, with only one resonant wavelength interaction occurring in each undulator module, radiation at the non-resonant wavelength will effectively propagate in free space. Two effects can be expected from this. Firstly, the non-resonant wavelength in an undulator module  will undergo free-space diffraction (unless it is an harmonic when some guiding can occur~\cite{psase}), which will reduce the net coupling of the radiation to the electron beam and so  increase the effective gain length of the interaction for both colours. Secondly, the addition of extra relative slippage between the radiation and the electrons in the non-resonant undulator modules may reduce the radiation bandwidth~\cite{ipac,wu,psase,hbsase,gainMod} and affect the spectrum by introducing modal effects~\cite{mlsase}, the latter being the subject of current study. 

\section{Simulation Results}

Simulations  using this modified version of Puffin are now presented that model a 2-colour FEL interaction using a single electron beam in sequential undulators tuned to two different resonant wavelengths. The first example is seeded by temporally coherent external laser sources at both of the resonant frequencies with intensities $\sim 10^{-4}$ less than the normal FEL saturation power. The seed powers are assumed constant over the duration of the electron pulse. These assumptions result in a simpler (cleaner)  electron beam phase space from which the nature of the 2-colour interaction is more easily observed. While such seeding is feasible at longer wavelengths, e.g. in the UV, it is not yet possible in the X-ray. Hence, the second example starts up from the electron beam shot-noise i.e. 2-color SASE to demonstrate X-ray operation. Radiation output spectra for differently tuned 2-colour FELs are then summarised. 

In a 2-color FEL, there will  be a small drift section between each undulator module that may require radiation/electron beam phase matching using small tuning chicanes. Although Puffin is capable of modelling the drift sections none are modelled here and an instantaneous change in undulator parameter is applied. Puffin is also used here in the 1D limit as described in \cite{POP}, so that simulations cost considerably less computational effort and the 2-colour interaction can be observed in its simplest form. More detailed effects including diffraction, that can be expected to alter the optimum set-up, will be investigated in future work.

\begin{figure}
\centering
\includegraphics[width=\columnwidth]{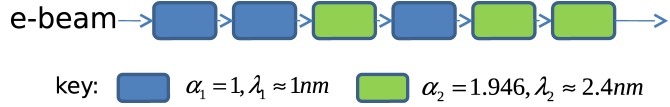}
\caption[Undulator layout]{Modular undulator layout for well contrasted 2-colour operation at  wavelengths $\lambda_1=1$nm and $\lambda_2=2.4$nm.}
\label{figure1}
\end{figure}

Modules with a fixed number of undulator periods are used which, while reflecting a probable experimental configuration, limits somewhat the options for driving both colours to the same powers. To compensate  for the difference in gain lengths of the alternating modules ($l_{g2} = \alpha^{-2/3}\l_{g1}$), the $N_m=6$ undulator modules are arranged in the configuration of Fig.~\ref{figure1}.  
This has a greater similarity to the configuration of~\cite{gainMod},  allowing both colours to experience gain periodically through the interaction, and is unlike the configurations of~\cite{slac,2colSACLA} which amplify each wavelength in two consecutive interactions.  

The  limit on the beam energy spread which must be satisfied for gain, $\Delta \gamma/\gamma_0 \lesssim \alpha^{2/3} \rho_1$,  also approximates the energy spread at FEL saturation.The undulator configuration of Fig.~\ref{figure1} has modules resonant at the wavelengths [1,1,2,1,2,2] and is chosen to avoid too much initial amplification of the longer wavelength. When the two colours are well contrasted, as in the following simulations, the longer wavelength interaction could inflate the beam energy spread to impede the interaction at the shorter wavelength, which has the tighter energy spread constraint.   For the same reason, it is preferable that the shorter wavelength is driven close to saturation prior to the longer and then ending the interaction at the longer wavelength.

The undulator module layouts in all of the following cases are designed with these arguments in mind, and are chosen to avoid full saturation of either colour. The post-saturation regime may be the subject of further publications.

\subsection{Seeded Example}
The parameters for the first seeded case are shown in Table \ref{table}. 
\begin{table}
\centering
\caption[Table of parameters]{Parameters for the 2-Colour Simulations}
\begin{tabular}{|  c   | c  |   c   |  c  |}
\hline
& Seeded & SASE & SASE scan \\
\hline
$\rho$ & $0.005$ & $0.005$ & $0.0011$ \\
\hline
$\gamma_0$ & $6316$ & $6316$ & $6900$ \\
\hline
$\sigma_\gamma/\gamma_0$ & $10^{-4}$ & $10^{-4}$ & $10^{-4}$ \\
\hline
$\bar{a}_{w0}$ & $1.0$ & $1.0$ & $1.2$ \\
\hline
$N_m$ & $6$ & $8$ & $10$ \\ 
\hline
$N_w$ & $25$ & $25$ & $100$ \\ 
\hline
$\alpha_1$ & $1$ & $1$ & $1$ \\ 
\hline
$\alpha_2$ & $1.946$ & $1.946$ & scan \\
\hline
\end{tabular}
\label{table}
\end{table}
A small seed field is injected at each resonant frequency, and the electron beam has a `flat-top' current profile. In the scaled frame $\bar{z}_2= (ct-z)/l_{c1}$, where $l_{c1}=\lambda_1/4\pi\rho_1$ is the cooperation length~\cite{nprev} of the shorter wavelength, the scaled resonant frequencies $\bar{\omega}=\omega/2\rho\omega_1$ are $\bar{\omega}_1 = 100$ and $\bar{\omega}_2 \approx 41.7$. In addition to the scaled radiation field $A_\perp$ of~\cite{POP}, it is useful to define a measure of the spectral bunching of the $N$ electrons as: 
\begin{align}
\tilde{b}(\bar{z},\bar{\omega}) = \frac{1}{\sqrt{2\pi}}\int_{-\infty}^{\infty} s\left(\bar{z},\bar{z}_{2}\right) e^{-i\bar{\omega} \bar{z}_2} d\bar{z}_2
\end{align}
where:
\begin{align}
	s\left(\bar{z},\bar{z}_2\right) = \frac{1}{\bar{n}_p} \sum_{j=1}^{N} e^{-i\bar{z} / 2 \rho} \delta(\bar{z}_{2} - \bar{z}_{2j}\left(\bar{z}\right)), \label{bunching}
\end{align}
is the source term of the unaveraged radiation wave equation in the Compton limit~\cite{POP} for the $\alpha_1$ undulator and $\bar{n}_p$ is the peak linear density of electrons per unit $\bar{z}_2$. 

\begin{figure}
\centering
\includegraphics[width=\columnwidth]{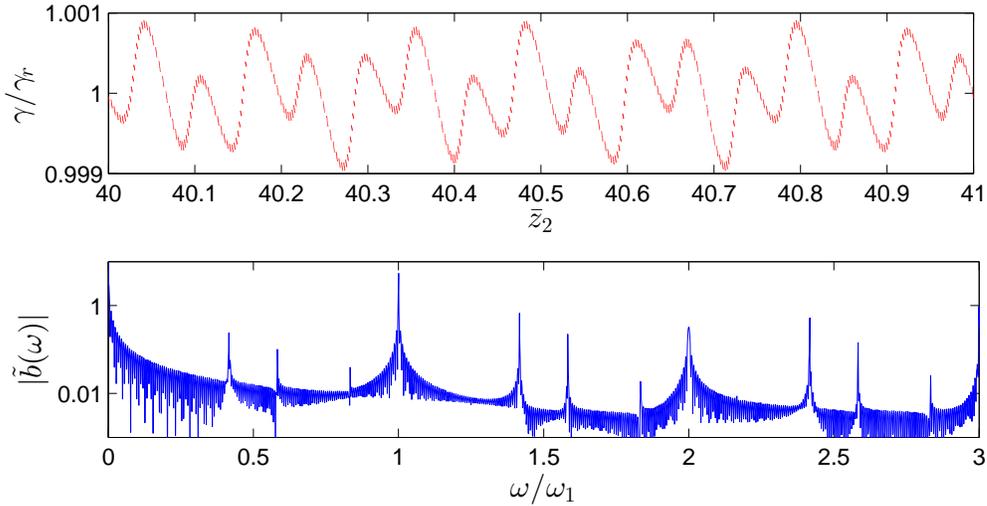}
\vspace*{-\baselineskip}
\caption[Bunching at end of sim]{Electron parameters at output of 3rd undulator module: Detail of electron phase space between $40<\bar{z}_2<41$ (top); and modulus of spectral bunching parameter $\tilde{b}(\bar{\omega})$ (bottom) at the output of the 3rd undulator module at  $\bar{z}= 4.71$.}
\label{figure2}
\end{figure}

Detail of the electron phase space and the modulus of the spectral bunching parameter is shown in Fig.\ \ref{figure2} at the end of the third module, and still in the linear regime,  where $\bar{z}\approx 4.71$. 
The bunching spectrum, shown in the bottom plot of Fig.\ \ref{figure2}, clearly shows bunching components at both frequencies, $\omega / \omega_1 = 1$ and $0.417$. It is interesting to note here that the beam also displays bunching components at the harmonics and the sums and differences between all of these frequencies. In the beam phase space of the top plot, these bunching components are seen in the superimposed periodic electron modulations of periods in $\bar{z}_2$ of $\lambda_{1,2}=4\pi\rho_1\times\omega_1 / \omega_{1,2} \approx 0.063$ and $0.15$ respectively.   

\begin{figure}
\centering
\includegraphics[width=\columnwidth]{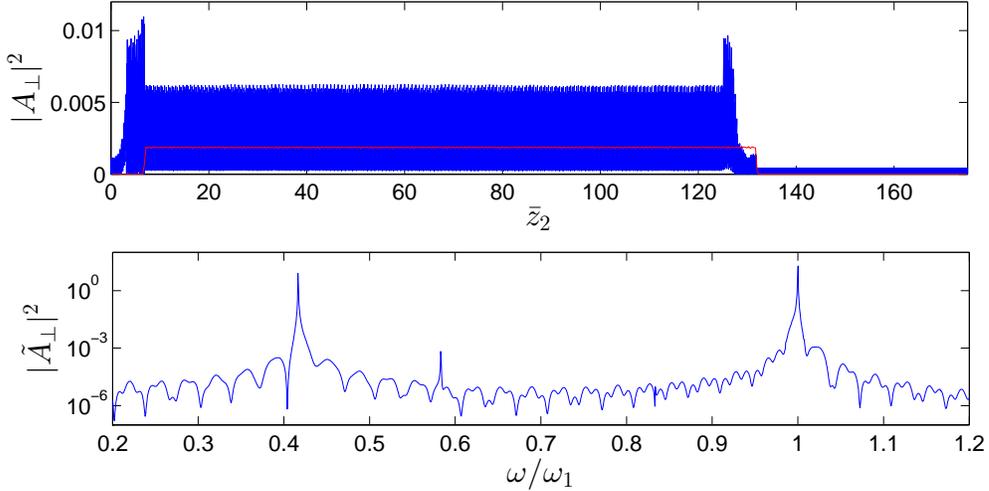}
\vspace*{-\baselineskip}
\caption[Bunching at end of sim]{Scaled radiation intensity $|A_\perp|^2$ as a function of the scaled temporal coordinate $\bar{z}_2$ (top) and the scaled spectral intensity $|\tilde{A}_\perp|^2$ (bottom) at the output of the 3rd undulator module at $\bar{z}= 4.71$. The red line in the temporal intensity plot indicates the current profile of the electron beam.}
\label{figure3}
\vspace*{-\baselineskip}
\end{figure}

The corresponding scaled `instantaneous' temporal intensity (i.e. including the fast oscillatory terms of the field~\cite{POP}) and the  scaled spectral intensity at the end of the third module are plotted in Fig.\ \ref{figure3}. 
The head of the electron pulse (shown in red in the top  window) starts the beginning of the interaction at $\bar{z}_2=0$ , and slips behind the radiation towards larger values of $\bar{z}_2=\bar{z}$ as the interaction progresses through the undulator. Coherent Spontaneous Emission at the tail and head of the rectangular current distribution are visible in the top intensity plot. Detail of the spectrum in the bottom plot shows the strong emission at the resonant frequencies $\omega / \omega_1 = 1$ and $0.417$, of the two undulator modules. A weaker emission is seen at the difference frequency $\omega / \omega_1 = 1-0.417=0.583$.



\begin{figure}
\centering
\includegraphics[width=\columnwidth]{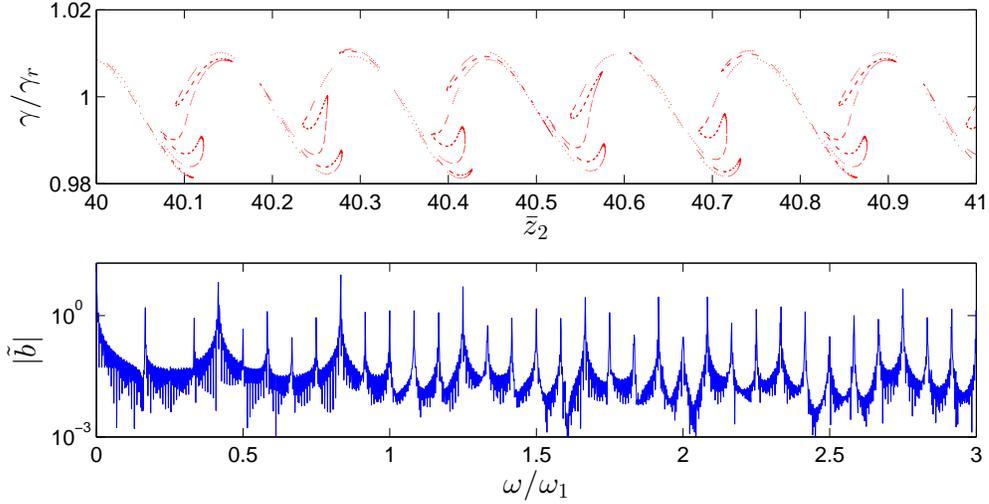}
\vspace*{-\baselineskip}
\caption[Bunching at end of sim]{As Fig.\ \ref{figure2} at output from 6th undulator module at $\bar{z}=9.42$.}
\label{figure4}
\end{figure}

\begin{figure}
\centering
\includegraphics[width=\columnwidth]{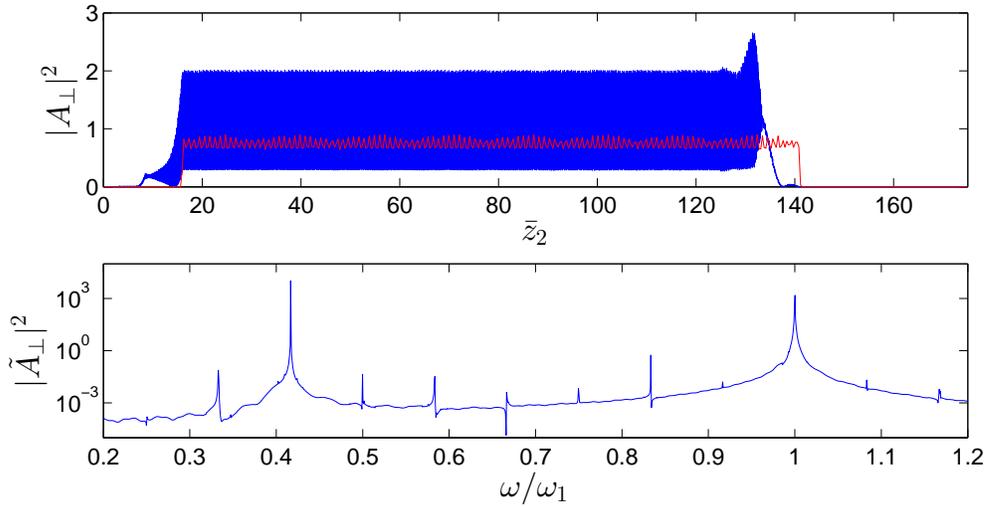}
\vspace*{-\baselineskip}
\caption[Bunching at end of sim]{As Fig.\ \ref{figure3} at output from 6th undulator module at $\bar{z}=9.42$.}
\label{figure5}
\end{figure}

Figs.\ \ref{figure4} and \ref{figure5} plot the same output at the end of the sixth undulator module at $\bar{z}=9.42$, where the longer wavelength is close to saturation, slightly ahead of the shorter.
A quite complex electron phase space has developed as a result of the 2-colour interaction which has generated spectral intensities that are approximately equal for the two colours. 


The interaction at the longer wavelength $\lambda_2$ has a greater energy spread tolerance, so the energy modulations at $\lambda_1$ do not have as great an effect on the $\lambda_2$ interaction. Due to dispersion in non-resonant undulators, this can reduce bunching at the shorter wavelength $\lambda_1$ in regions where the energy modulation at $\lambda_2$  shifts the beam off-resonance. The resultant oscillation of the  bunching at $\lambda_1$ therefore occurs with period $\lambda_2$, and consequently bunching is observed at the sum and difference frequencies $\omega_1 + \omega_2 $ and $\omega_1 - \omega_2$.

This effect generates more frequency components in the bunching as the FEL process develops. As seen in Fig. \ref{figure4}, the same mechanism applies as the bunching at $\lambda_2$ and its harmonics grow, so that components develop not just at the sum and difference of the lasing frequencies, but also with the higher harmonic frequencies.

\subsection{2-Color SASE Example}
The 2-colour FEL interaction is now modelled starting from noise via SASE~\cite{bnp}.  A similar, but extended, undulator beamline as that of Fig.\ \ref{figure1} is used with $N_m=9$ modules in a $[1,1,2,1,1,2,1,2,2]$  configuration, to allow for a longer  build-up from noise. Otherwise all parameters are identical to the previous seeded case. 


One factor that is present in the 2-colour SASE, that does not exist in the seeded case, is the potential for phase mis-matching between the radiation and the electron beam due to the extra induced slippage in the alternate non-resonant undulator modules. This may be mitigated by using modules $\lesssim 1$ gain length, ensuring relative slippages are not significantly longer than a cooperation length, so maintaining beam/radiation phase matching. While additional slippage may also assist in improving the coherence of the radiation field as discussed above, this should not significantly affect the results presented here. 

\begin{figure}[t]
\centering
\includegraphics[width=\columnwidth]{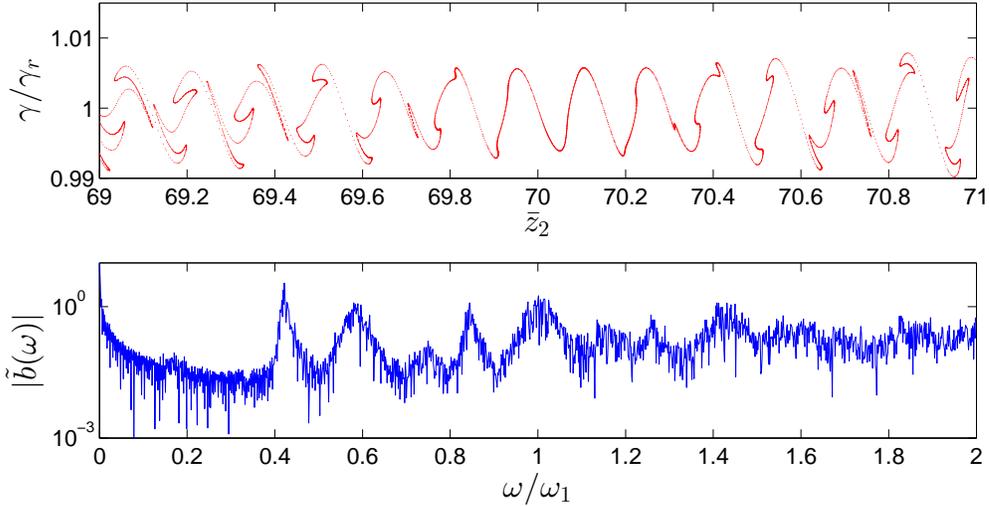}
\caption[Bunching at end of sim]{Electron parameters at output of $9^{th}$  undulator module for SASE: Detail of electron phase space between $69<\bar{z}_2<71$ (top); and modulus of spectral bunching parameter $\tilde{b}(\bar{\omega})$ (bottom) at the output of the $9^{th}$  undulator module at $\bar{z}= 14.12$.}
\label{figure6}
\end{figure}

\begin{figure}
\centering
\includegraphics[width=\columnwidth]{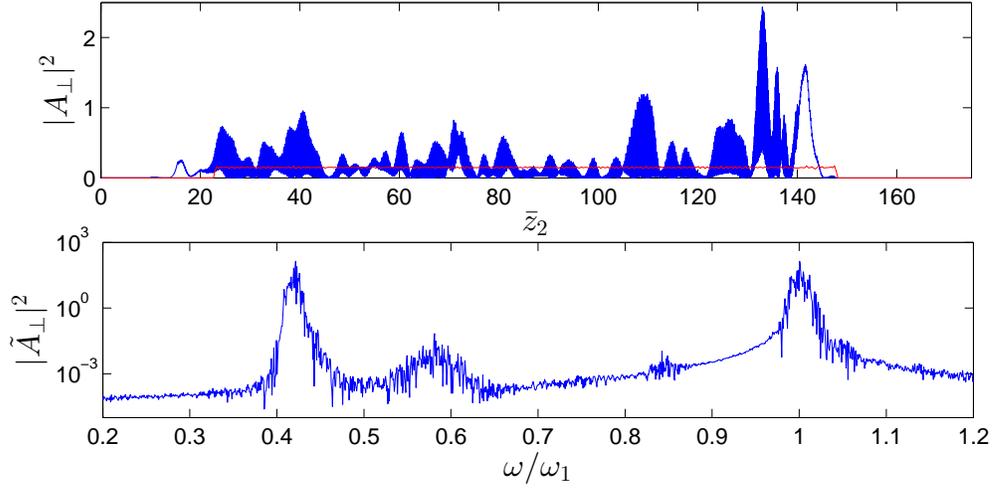}
\caption[Radiation at end of sim]{Scaled radiation SASE intensity $|A_\perp|^2$ as a function of the scaled temporal coordinate $\bar{z}_2$ (top) and the scaled spectral intensity $|\tilde{A}_\perp|^2$ (bottom) at the output of the $9^{th}$ undulator module at $\bar{z}= 14.12$. The red line in the temporal intensity plot indicates the current profile of the electron beam.}
\label{figure7}
\end{figure}

As the system now starts from noise, the resultant beam phase space is less structured but maintains similar characteristics to the previous seeded simulations. Similarly to Fig.~\ref{figure4}, the electron phase space close to saturation is seen from Fig.~\ref{figure6} to have bunching components at both resonant frequencies $\omega_{1,2}$ and at the sum and difference frequencies. 
The corresponding scaled radiation  intensity and spectrum equivalent to the seeded case of Fig.~\ref{figure5}  are plotted in Fig. \ref{figure7}, where the spectral intensity levels at both resonant frequencies are seen to be approximately equal. Unlike the seeded case, the temporal intensity has the familiar chaotic pulse structure. 

\begin{figure}
\centering
\includegraphics[width=\columnwidth]{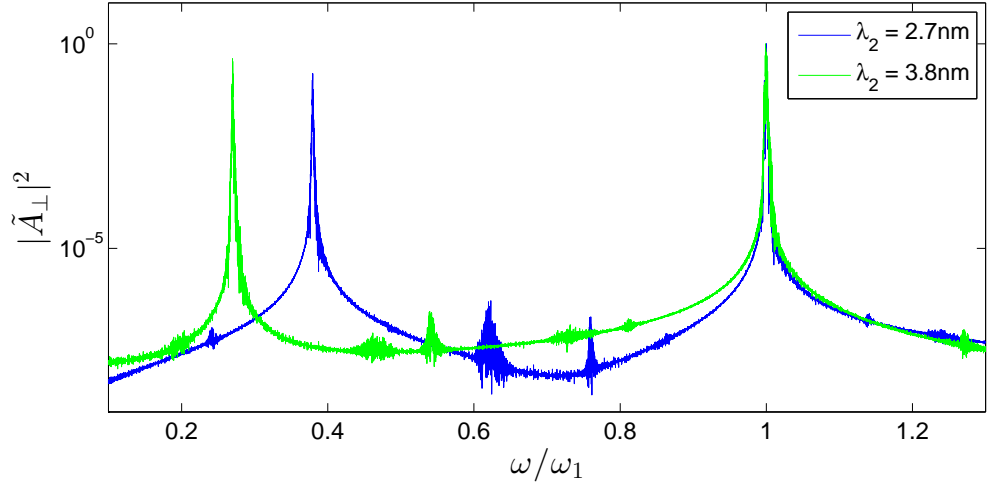}
\caption[Bunching at end of sim]{Scaled radiation intensity spectra at output for two different 2-colour operating frequencies}
\label{figure8}
\end{figure}

Similar 2-colour SASE simulations were also carried out for a system with a longer gain length more typical of an X-ray FEL, and with the electron beam and undulator parameters similar to the Athos FEL proposed at SwissFEL~\cite{swissfel}. The FEL parameter is  reduced to $\rho_1 = 0.001$, and the number of undulator periods per module has been increased to $N_w = 100$ to accomodate the longer gain length. There are now $N_m = 10$ modules in a $[1,1,2,1,1,2,1,1,2,2]$ configuration. Maintaining the same higher resonant wavelength of $\lambda_1=1$nm for $\alpha_1=1$, simulations for two different 2-colour configurations  were carried out for $\lambda_2=2.7$nm and $\lambda_2=3.8$nm corresponding to $\alpha_2=1.95$ and $2.37$ respectively. The normalised saturated SASE spectra (i.e. neglecting the strong CSE emitted  from the head and tail of the flat-top electron beam) of the scaled output intensities are plotted in Fig. \ref{figure8} and demonstrates  that the principle of such 2-colour FEL operation applies across a wide range of wavelengths and to similar saturated radiation powers. No attempt has been  made to optimise the output here, so that the output at a given wavelength may be improved upon. The statistical nature of the output is also unaccounted for in these single-shot simulations. 


\section{Conclusion}
The principle of 2-colour FEL amplifier lasing using a mono-energetic electron beam in a series of alternately tuned undulator modules was, for the first time, simulated at widely spaced, non-harmonic wavelengths. Such simulations are only practical using non-averaged, broad bandwidth FEL simulation codes such as Puffin~\cite{POP}.  To enable 2-colour modelling, Puffin was modified from that described in~\cite{POP}  to allow for any general variation in the magnetic field strength along the undulator .  

Both seeded and SASE 2-colour operation were demonstrated to saturation and gave similar peak spectral intensities at both the resonant wavelengths. Such simultaneous FEL lasing to saturation by one electron beam at two distinct, harmonically uncoupled, wavelengths is perhaps not an immediately intuitive result. However, the electron phase-space evolution has a rich structure, including bunching at the sum and difference frequencies and  harmonic components, that clearly demonstrates the simultaneous electron bunching and emission at the two distinct wavelengths. This result may open up other avenues for further research, for example in the generation of higher-modal radiation emission, multi-modal emission using multiply tuned undulator modules, or perhaps non-linear frequency mixing processes in the FEL~\cite{3beam2}. 

In the case of the 2-color SASE simulations, only a limited number of simulations were performed for each case so that the statistical nature of the 2-colour processes were not explored. In both the seeded and SASE cases the scaled saturation variables indicates that parameters such as photon numbers and pulse durations will be similar to those for the equivalent single colour interaction.  

Other factors must be taken into account in future 3D simulations. Diffraction will clearly have an effect during `free' propagation of the radiation in the non-resonant undulator modules. Another factor is the transverse beam matching to different undulator modules. This could become more problematic for larger differences in $\bar{a}_w$ between undulator modules and may limit the range of the 2-colour operation. Clearly, further examination in $3$D is needed to identify and perhaps find solutions to such issues.

\ack
We gratefully acknowledge the computing time granted by the John von Neumann Institute for Computing (NIC) and provided on the supercomputer JUROPA at Jülich Supercomputing Centre (JSC), under project HHH20; STFC Memorandum of Agreement No. 4132361; and ARCHIE-WeSt High Performance Computer, EPSRC grant no. EP/K000586/1 \\

\end{document}